
\magnification=1200
\baselineskip=24truept
\hsize 14.9truecm \hoffset 1.2truecm
\vsize 22.2truecm\voffset .2truecm
\font \titolo=cmbx12 scaled\magstep0

\font \ninerm=cmr10 at 10truept
\font \ninei=cmti10 at 10truept
\outer\def\beginsection#1\par{\medbreak\bigskip
      \message{#1}\leftline{#1}\nobreak\medskip\vskip-\parskip}
\def \TeX{T\kern-.1667em\lower.5ex\hbox{E}\kern-.125em X}
\def \Sg {\Sigma}

\def \om {\omega}

\def \th {\theta}

\def \La {\Lambda}

\def \a {\alpha}

\def \ga {\gamma}

\def \da {\delta}
\def \ep {\epsilon}
\def \part {\partial}

\def \um {{1\over 2}}

\def \Tr {\hbox{Tr}}

\def \sqr#1#2{{\vcenter{\hrule height.#2pt
       \hbox{\vrule width.#2pt height#1pt \kern#1pt
          \vrule width.#2pt}
       \hrule height.#2pt}}}

\def\lsim{\mathrel{\rlap{\lower4pt\hbox{\hskip1pt$\sim$}}
    \raise1pt\hbox{$<$}}}         
\def\gsim{\mathrel{\rlap{\lower4pt\hbox{\hskip1pt$\sim$}}
    \raise1pt\hbox{$>$}}}         

\def\IR{{\hbox{{\rm I}\kern-.2em\hbox{\rm R}}}}
\def\IH{{\hbox{{\rm I}\kern-.2em\hbox{\rm H}}}}
\def\IC{{\ \hbox{{\rm I}\kern-.6em\hbox{\bf C}}}}
\def\IZ{{\hbox{{\rm Z}\kern-.4em\hbox{\rm Z}}}}
\rightline{DFTT 67/93}
\rightline{UCD-93-33}
\rightline{gr-qc/9311007}
\vskip 3ex
\centerline{\titolo Equivalent Quantisations
of (2+1)-Dimensional Gravity%
\footnote{\raise2pt \hbox{\ninerm*}}{\ninerm Work supported in part by
U.S.~Department of Energy Grant DE-FG03-91ER40674 and INFN Iniziativa
Specifica TO10~(FI2).}}
\vskip 0.7truecm
\centerline{S.~Carlip }
\centerline{\it Department of Physics, University of California}
\centerline{\it Davis, CA 95616, USA}
\centerline{\ninei (email:~carlip@dirac.ucdavis.edu,
 telefax:~01-916-752-4717)}
\vskip 0.7truecm
\centerline{J.E.~Nelson}
\centerline{\it Dipartimento di Fisica Teorica dell'Universit\`a di Torino}
\centerline{\it via Pietro Giuria 1, 10125 Torino, Italy}
\centerline{\ninei (email:~nelson@to.infn.it, telefax:~039-11-6707214)}
\vskip 1ex
\leftline {\ninerm P.A.C.S.\ 04.60}
\vskip 1ex

\centerline{\bf Abstract}

For spacetimes with the topology $\IR\!\times\!T^2$, the action of
(2+1)-dimensional gravity with negative cosmological constant $\La$ is
written uniquely in terms of the time-independent traces of
holonomies around two intersecting noncontractible paths on $T^2$.
The holonomy parameters are related to the moduli on slices of constant
mean curvature by a time-dependent canonical transformation which
introduces an effective Hamiltonian.  The quantisation
of the two classically equivalent formulations differs by terms of
order $O(\hbar^3)$, negligible for small $|\La|$.
\vfill\eject
\centerline{\bf Classical Theory}
\vskip .5ex

Amongst the various approaches to (2+1)-dimensional quantum gravity
without matter couplings, there are two that have seen substantial
development in the last few years: the second-order metric
formalism of Moncrief [1] and the first-order holonomy algebra
formalism of Nelson and Regge [2], inspired by Witten [3].  Both
start with a spacetime with the topology $\IR\!\times\!\Sg$, where
$\Sg$ is a closed Riemann surface.  In the first-order formalism,
the variables are two sets of $6g-6$ ($g>1$) time-independent traces
of holonomies around intersecting noncontractible paths on $\Sg$.
Their Hamiltonian is zero, since the constraints have been solved
exactly. In the second-order formalism the constraints are also
solved exactly, but the remaining variables are $6g-6$ time-dependent
coordinates for the Teichm\"uller space of $\Sg$ and their conjugate
momenta, with a time development generated by a nontrivial effective
Hamiltonian.

When the cosmological constant is zero and $\Sigma$ is a torus $T^2$
(for which there are only two degrees of freedom), it has been
shown that the two approaches are classically equivalent,
and are related through a time-dependent canonical transformation.
The corresponding quantum theories are closely related, but are not
quite equivalent in the most natural operator ordering [4].
The purpose of this letter is to affirm that the same is true when
the cosmological constant is negative, and that the unique action
of (2+1)-dimensional gravity can be written in the equivalent forms
$$\eqalignno{\int\!dt \int\!d^2x\,\pi^{ij} {\dot g}_{ij}
= & \int\!dt\int\!d^2x\,2\ep^{ij}\ep_{abc}\,e^c_j\,{\dot{\om}}^{ab}_i \cr
= &  \int\um (\bar p dm + p d\bar m) +
H d\tau -d(p^1m_1 +p^2m_2) &(1)\cr = & \int \a(r_1^-dr_2^- - r_1^+ dr_2^+)
\cr}$$
In (1), $m= m_1+im_2$ is the complex modulus of the torus, related to the
spatial components $g_{ij}$ of the metric tensor on a slice of constant
mean curvature by
$$g^{-1/2} g_{ij}={1\over m_2}\left(\matrix{m_1^2+m_2^2&m_1\cr
m_1&1 \cr}\right) \eqno(2)$$
and $p=p^1+ip^2$ is the conjugate momentum, defined by [1]
$$p^a= (\um g_{ij}\pi-\pi_{ij}){\partial\ \over {\partial {m_a}}}g^{ij}
\qquad (a=1,2) \eqno(3)$$
The $r_{1,2}^{\pm}$ are the variables of the SL(2,$\IR$) holonomies [2],
expressed as
$$\eqalign{R_1^{\pm}&= \cosh{r_1^{\pm} \over 2}\cr
R_2^{\pm}&= \cosh{r_2^{\pm} \over 2}\cr}\eqno(4)$$
where the subscripts 1 and 2 in (4) refer to two intersecting paths
$\ga_1, \ga_2$ on $\Sg$ with intersection number $+1$. (A third holonomy,
$R^\pm_{12} = \cosh{(r_1^{\pm} + r_2^{\pm})/2}$, corresponds to the
path $\ga_1\cdot\ga_2$, which has intersection number $-1$ with $\ga_1$
and $+1$ with $\ga_2$.)

To arrive at (1), the first step is to choose spatial hypersurfaces $\Sg$
labelled by $\Tr K=-g^{-1/2} \pi = \tau=const.$ and to solve the
constraints, which for negative cosmological constant $\La=-{1/{\a^2}}$
read
$$\eqalign{R^{ab} &=d\omega^{ab} -\omega^{ac}\wedge
  \omega_{c}{}^{b}= {1\over {\a^2}}e^a\wedge e^b \cr
R^{a} &= de^a -\om^{{ab}}\wedge e_b =0\cr}\eqno(5)$$
with $a,b,c=0,1,2$.
In the time gauge $e^0_i=0$, these can be satisfied by the choice
$$\eqalignno{&e^0 =dt \cr
&e^1= {\a \over 2} \left[(r_1^+ - r_1^-)dx +(r_2^+ - r_2^-)dy\right]
 \sin{t \over {\a}} &(6a)\cr
&e^2= {\a \over 2} \left[(r_1^+ + r_1^-)dx +(r_2^+ + r_2^-)dy\right]
 \cos{t \over {\a}} \cr
&\om^{12}=0  &\cr
&\om^{01}= -{1 \over 2} \left[(r_1^+ - r_1^-)dx +(r_2^+ - r_2^-)dy\right]
 \cos{t \over {\a}} &(6b)\cr
&\om^{02}= {1 \over 2} \left[(r_1^+ + r_1^-)dx +(r_2^+ + r_2^-)dy\right]
 \sin{t \over {\a}} \cr} $$
where $x$ and $y$ have period one and the time coordinate $t$ is
determined by $\tau=-{2 \over {\a}}\cot {2t \over {\a}}$.
The components (6a--b) can be used in two ways.  Firstly, we
calculate the metric tensor $g_{ij}=e^a{}_i e^b{}_j\eta_{ab}$
of $\Sg$, and thence from (2) and (3) the
modulus\footnote{\raise1pt \hbox{\ninerm*}}{\ninerm This result is
essentially equivalent to that of Fujiwara and Soda [1], although
some care must be taken in the conversion; in particular, as they
note in section 5, their spatial coordinates need not have period one.}
$$m= \left(r_1^-e^{it/\a} + r_1^+e^{-{it/\a}}\right)
 \left(r_2^-e^{it/\a} + r_2^+e^{-{it/\a}}\right)^{\lower2pt%
 \hbox{$\scriptstyle -1$}} \eqno(7)$$
and its conjugate momentum
$$p= -{i\a\over 2\sin{2t\over \a}}\left(r_2^+e^{it/\a}
 + r_2^-e^{-{it/\a}}\right)^{\lower2pt%
 \hbox{$\scriptstyle 2$}} \eqno(8)$$
These variables satisfy the Poisson bracket algebra
$$(\bar m,p)=(m,\bar p)=-2, \qquad (m,p)=(\bar m,\bar p)=0 \eqno(9)$$
Secondly, we compute the traces of the SL(2,$\IR$) holonomies
corresponding to the two generators of the fundamental group $\pi_1(\Sg)$,
using the decomposition of the spinor group of $\hbox{SO}(2,2)$
as a tensor product $\hbox{SL(2,$\IR$)}\otimes\hbox{SL(2,$\IR$)}$. This
approach recovers the traces (4), which satisfy the nonlinear
classical Poisson bracket algebra [2]
$$(R_1^{\pm},R_2^{\pm})=\mp{1\over 4\a}(R_{12}^{\pm}-
 R_1^{\pm}R_2^{\pm}) \eqno(10)$$
consistent with the brackets
$$(e^a_i({\bf x}),\om^{bc}_j({\bf y}))=- \um \ep_{ij}\ep^{abc}
\da^2({\bf x-y}) \eqno(11)$$
obtained from (1). Equation (10) implies that the holonomy parameters
satisfy
$$(r_1^\pm,r_2^\pm)=\mp {1\over\a}, \qquad (r^+,r^-)=0 \eqno(12)$$
It is easily checked that the brackets (12) induce (9) and (10).  The
holonomy parameters $r_{1,2}^{\pm}$ are thus related to the modulus $m$
and momentum $p$ through a (time-dependent) canonical transformation, as
expressed in (1).

The Hamiltonian in equation (1) now takes the form
$$H=g^{1/2}={\a^2\over 4}\sin{2t\over \a}\,(r_1^-r_2^+ - r_1^+r_2^-)
\eqno(13)$$
which generates the development of the modulus (7) and momentum (8)
through
$${dp\over d\tau}=(p,H),\qquad {dm\over d\tau}=(m,H) \eqno(14)$$
Alternatively, the Hamiltonian
$$ H^\prime={d\tau\over dt}H={4\over {\a^2}}\csc^2 {2t\over \a}H
\eqno(15)$$
generates evolution in coordinate time $t$ by
$${dp\over dt}=(p, H^\prime),\qquad {dm\over dt}=(m, H^\prime) \eqno(16)$$
The time-dependent moduli $m_1,m_2$  lie on a semicircle,
$$\left(m_1 - {{r_1^+ r_2^+ - r_1^- r_2^-}\over
{{r_2^+}^2 - {r_2^-}^2}}\right)^{\lower2pt\hbox{$\scriptstyle 2$}}
+ m_2^2 = \left({{r_1^+ r_2^- - r_1^- r_2^+}\over
{{r_2^+}^2 - {r_2^-}^2}}\right)^{\lower2pt\hbox{$\scriptstyle 2$}}
\eqno(17)$$
as can be seen from (7), again agreeing with Fujiwara and Soda [1].

The standard action of the modular group on the torus modulus,
$$\eqalign{&S: m\rightarrow
  -{1\over m},\qquad p\rightarrow{\bar m}^2 p\cr
  &T: m\rightarrow m+1,\qquad p\rightarrow p\cr}\eqno(18)$$
preserves the brackets (9).  The same group acts on the holonomy parameters
as
$$\eqalign{&S:r_1^{\pm}\rightarrow r_2^{\pm},\qquad
    r_2^{\pm}\rightarrow - r_1^{\pm}\cr
&T:r_1^{\pm}\rightarrow r_1^{\pm} + r_2^{\pm},\qquad
    r_2^{\pm}\rightarrow r_2^{\pm} ,\cr}\eqno(19)$$
preserving the brackets (12) and the Hamiltonians (13) and (15). On the
traces (4), the group action is
$$\eqalign{&S:R_1^{\pm}\rightarrow R_2^{\pm},
  \quad R_2^{\pm}\rightarrow R_1^{\pm},
  \quad R_{12}^{\pm}\rightarrow 2R_1^{\pm} R_2^{\pm} - R_{12}^{\pm}\cr
&T:R_1^{\pm}\rightarrow R_{12}^{\pm},\quad R_2^{\pm}\rightarrow R_2^{\pm},
\quad R_{12}^{\pm}\rightarrow 2R_{12}^{\pm} R_2^{\pm} - R_1^{\pm}\cr}
\eqno(20)$$
corresponding to the intersection number preserving exchanges
$$\eqalign{&S:\ga_1\rightarrow \ga_2^{-1},\qquad \ga_2\rightarrow
\ga_1\cr
&T:\ga_1\rightarrow \ga_1\cdot\ga_2 ,\qquad \ga_2\rightarrow \ga_2\cr}
\eqno(21)$$
It is easy to verify from (7--8) that the transformations (19) of $r^\pm$
induce the correct transformations (18) of $m$ and $p$.

The relationship between the metric and the holonomies can be further
checked by means of a quotient space construction.  The holonomies (4)
generate a subgroup
$$\Gamma =
\left\langle R_1^+\otimes R_1^-,R_2^+\otimes R_2^-\right\rangle \eqno(22)$$
of $\hbox{SL(2,$\IR$)}\otimes\hbox{SL(2,$\IR$)}$, which acts on
three-dimensional anti-de Sitter space as a group of isometries.  It
may be shown that the quotient of anti-de Sitter space by this group
is a spacetime with topology $\IR\!\times\!T^2$ whose induced metric
is precisely the metric $g_{\mu\nu} = \eta_{ab}e^a{}_\mu e^b{}_\nu$
obtained from the triads (6a).  The holonomies thus describe a geometric
structure of the type discussed by Mess [5].
\vskip 1ex

\centerline {\bf Quantisation}
\vskip .5ex

It is easy to quantise the system described so far. The quantisation of
(10) gives the weighted algebra [2]
$$R_1^{\pm}R_2^{\pm}e^{\pm i \th} - R_2^{\pm}R_1^{\pm} e^{\mp i \th}=
\pm 2i\sin\th\, R_{12}^{\pm} \eqno(23)$$
with $\tan\th= - {\hbar/8\a}$, consistent with the commutators
$$[r_1^{\pm}, r_2^{\pm}] = \pm 8i\th,\qquad[r^{\pm},r^{\mp}] = 0\eqno(24)$$
when the holonomies (4) are represented by the operators
$$R_a^{\pm} = \sec\th \cosh{r_a^{\pm} \over 2} \qquad (a=1,2) \eqno(25)$$
With the ordering of (7--8) and (13) it follows that
$$\eqalign{[\bar m,p]=[m,\bar p]= 16i\a\th, \qquad
  [m,p] = [\bar m,\bar p] = 0\cr
[p, H^\prime] = -8i\a\th {dp\over dt}, \qquad  [m, H^\prime] =
-8i\a\th {dm\over dt}\cr}\eqno(26)$$
which, with $\th = - \tan^{-1}({\hbar/{8\a}})$, differ from the
direct quantisation of (12), (9) and (16)
by terms of order $O(\hbar^3)$, small when $|\La|= 1/\a^2$ is small.

Full details, including the action of the quantum modular group,
the precise relationship with Fujiwara and Soda's results [1], the
quantum group construction of reference [2], and the limit
$\La\rightarrow 0$, will be discussed elsewhere [6].

\vskip .7truecm

\centerline {\bf References}

\vskip 1ex
\item{[1]\ }V.~Moncrief, J.\ Math.\ Phys.\ {\bf 30} (1989) 2907;
 A.~Hosoya and K.~Nakao, Class.\ Quantum Grav.\ {\bf 7} (1990) 163,
 Prog.\ Theor.\ Phys. {\bf 84} (1990) 739; Y.~Fujiwara and J.~Soda,
 Prog.\ Theor.\ Phys. {\bf 83} (1990) 733.
\item{[2]\ }J.~E.~Nelson and T.~Regge, Phys.\ Lett.\ {\bf B272} (1991)
 213, Nucl.\ Phys.\ {\bf B328} (1989), Commun.\ Math.\ Phys.\ {\bf 141}
 (1991) 211, Commun.\ Math.\ Phys.\ {\bf 155} (1993) 561; J.~E.~Nelson,
 T.~Regge and F.~Zertuche, Nucl.\ Phys.\ {\bf B339} (1990) 516.
\item{[3]\ }E.~Witten, Nucl.\ Phys.\ {\bf B311} (1988/89) 46-78.
\item{[4]\ }S.~Carlip, Phys.\ Rev.\ {\bf D42} (1990) 2647, Phys.\ Rev.\
 {\bf D45} (1992) 3584, Phys.\ Rev.\ {\bf D47} (1993) 4520.
\item{[5]\ }G.~Mess, ``Lorentz Spacetimes of Constant Curvature,''
 Institut des Hautes Etudes Scientifiques preprint IHES/M/90/28 (1990).
\item{[6]\ }S.~Carlip and J.~E.~Nelson, in preparation.

\bye